\documentclass{aastex631}
\usepackage{epstopdf}
\usepackage{natbib}
\usepackage{graphicx}
\usepackage{CJK}

\usepackage[percent]{overpic}
\usepackage{times}
\usepackage{graphicx}
\usepackage{amsmath}

\usepackage{subfigure}
\usepackage{natbib}
\usepackage{txfonts}
\usepackage{journals}
\usepackage{xcolor}
\usepackage[utf8]{inputenc}
\usepackage[overload]{empheq}
\usepackage{verbatim}

\usepackage{booktabs}

\newcommand{\uvwa}{$uvw1$}
\newcommand{\uvmb}{$uvm2$}
\newcommand{\uvwb}{$uvw2$}
\newcommand{\bp}{$G_{\rm BP}$}
\newcommand{\rp}{$G_{\rm RP}$}
\newcommand{\ebv}{$E(B-V)$}
\newcommand{\teff}{$T_{\rm eff}$}
\newcommand{\feh}{$\rm [Fe/H]$}
\newcommand{\logg}{${\rm log}\,g$}
\newcommand{\av}{$A_{\rm V}$}
\newcommand{\rv}{$R_{\rm V}$}
\newcommand{\ebr}{$E_{G_{\rm BP},G_{\rm RP}}$}
\newcommand{\enb}{$E_{{\rm NUV},G_{\rm RP}}$}
\newcommand{\efb}{$E_{{\rm FUV},G_{\rm RP}}$}
\newcommand{\ewab}{$E_{{uvw1},G_{\rm RP}}$}
\newcommand{\embb}{$E_{{uvm2},G_{\rm RP}}$}
\newcommand{\ewbb}{$E_{{uvw2},G_{\rm RP}}$}
\newcommand{\ks}{$K_{\rm S}$}

\shortauthors{}

\submitjournal{ApJS}
\begin{document}
\begin{CJK*}{UTF8}{gbsn}

\title{Massive Acquisition of Ultraviolet Color Excess Information from GALEX and UVOT Bands}
\author[0009-0008-6604-2517]{Dongliang Yang (杨栋梁)}
\affiliation{Department of Physics,
               Hebei Key Laboratory of Photophysics Research and Application, 
               Hebei Normal University,
               Shijiazhuang 050024, P.\,R.\,China}
\author[0000-0002-2473-9948]{Mingxu Sun (孙明旭)}
\affiliation{Department of Physics,
               Hebei Key Laboratory of Photophysics Research and Application, 
               Hebei Normal University,
               Shijiazhuang 050024, P.\,R.\,China}
\author[0000-0003-3168-2617]{Biwei Jiang(姜碧沩)}
\affiliation{School of Physics and Astronomy,
               Beijing Normal University,
               Beijing 100875, P.\,R.\,China}
\author[0000-0003-1359-9908]{Wenyuan Cui(崔文元)}
\affiliation{Department of Physics,
               Hebei Key Laboratory of Photophysics Research and Application,
               Hebei Normal University,
               Shijiazhuang 050024, P.\,R.\,China}
\author[0000-0003-1863-1268]{Ruoyi Zhang(张若羿)}
\affiliation{School of Physics and Astronomy,
               Beijing Normal University,
               Beijing 100875, P.\,R.\,China}
\author{Luyao Shi (史璐瑶)}
\affiliation{Department of Physics,
               Hebei Key Laboratory of Photophysics Research and Application, 
               Hebei Normal University,
               Shijiazhuang 050024, P.\,R.\,China}
\author{Jiachen Wei (魏嘉辰)}
\affiliation{Department of Physics,
               Hebei Key Laboratory of Photophysics Research and Application, 
               Hebei Normal University,
               Shijiazhuang 050024, P.\,R.\,China}
\correspondingauthor{Mingxu Sun; Wenyuan Cui}
\email{mxsun@hebtu.edu.cn; cuiwenyuan@hebtu.edu.cn}

\date{Accepted xxx. Received xxx; in original form xxx}

\begin{abstract}

This study employs stellar parameters from spectroscopic surveys and Zhang et al. based on Gaia XP spectra, along with photometric data from GALEX, UVOT, and Gaia, to obtain extensive ultraviolet color excess information for the relevant bands of GALEX and UVOT. By considering the impact of stellar parameters (\teff, \feh, and \logg) on intrinsic color indices, and utilizing the blue-edge method combined with a random forest algorithm, an empirical relationship between stellar parameters and intrinsic ultraviolet color indices is established. By combining observed color indices, the study derives color excesses for 11,624,802 and 65,531 stars in the GALEX/near-UV and far-UV bands, and 336,633, 137,739, and 253,271 stars in the UVOT/\uvwa, \uvmb, and \uvwb\ bands, constructing corresponding ultraviolet extinction maps. Notably, the color excess data for the GALEX/near-UV band shows a tenfold increase from previous results, with the extinction map covering approximately two-thirds of the sky at a resolution of 0.4$^{\circ}$. The typical uncertainties in the ultraviolet color excesses are 0.21 mag, 0.30 mag, 0.19 mag, 0.24 mag, and 0.21 mag for \enb, \efb, \ewab, \embb, and \ewbb, respectively. By comparing the spatial distributions of \rv\ derived from ultraviolet and Gaia optical band measurements with those obtained from infrared and optical data in previous works, it is evident that the \rv\ distributions based on the ultraviolet data show noticeable differences, with some regions even exhibiting opposite trends. This suggests that a single-parameter \rv\ extinction law may not be sufficient to simultaneously characterize the extinction behavior across infrared, optical, and ultraviolet bands.

\end{abstract}


\section{Introduction} 
\label{sec:intro}
Interstellar dust, although constituting only about 1\% of the interstellar medium, is the primary factor responsible for extinction \citep{1978ppim.book.....S}. The study of extinction is crucial for two reasons: first, since interstellar dust absorbs and scatters light, altering the brightness and color of stars, understanding extinction helps us accurately restore the intrinsic brightness and color of these objects. Second, constructing dust models based on observations allows us to glean important physical and chemical information about the dust itself. Interstellar extinction varies not only along different lines of sight and distances but also with environmental changes \citep{2003ApJ...598.1017D}. Different wavelengths exhibit varying sensitivities to extinction, with ultraviolet (UV) wavelengths being more sensitive compared to optical and infrared (IR) wavelengths. When \rv\ = 3.1, UV extinction is approximately three times that of optical extinction \citep{1990ARA&A..28...37M}. Obtaining UV extinction measurements offers a distinct advantage in regions of high Galactic latitude and relatively low extinction.

Observations in the UV bands are challenging due to the Earth's atmosphere blocking most UV radiation, which makes the use of space telescopes essential. The launch of the Galaxy Evolution Explorer (GALEX; \citealt{2014Ap&SS.354..103B}) provided a major advancement for UV observations. GALEX has generated data for nearly 80 million sources in the NUV band ($\lambda_{\rm eff}\sim2310\ \AA$) and about one-tenth as many sources in the FUV band ($\lambda_{\rm eff}\sim1528\ \AA$). Additionally, the UV/Optical Telescope (UVOT; \citealt{2005SSRv..120...95R}) has observed over 10 million sources, delivering UV photometric measurements across three bands: \uvwa\ (260.0 nm), \uvmb\ (224.6 nm), and \uvwb\ (192.8 nm), covering the region of the 2175 \AA\ extinction feature. These extensive UV surveys provide a comprehensive dataset that forms a solid basis for studying UV extinction.

Extinction maps are crucial for extinction correction and understanding the distribution of dust. \citet{1982AJ.....87.1165B} created a reddening map of the Milky Way using the HI/galaxy counts (HI/GC; \citealt{1978ApJ...225...40B}) method. \citet{1998ApJ...500..525S} (SFD98) constructed a 2D reddening map based on the far-infrared emission of dust by combining 100, 140, and 240 $\mu$m data from COBE/DIRBE and IRAS/ISSA, significantly improving the accuracy over the results of \citet{1982AJ.....87.1165B}. \citet{2010ApJ...719..415P} used a technique called the ``standard crayon" method, employing passively evolving galaxies as color standards to correct for the underestimation of dust in low dust temperature regions in the SFD98 dust map. In terms of 3D extinction maps, studies such as \citet{2014MNRAS.443.1192C} and \citet{2015ApJ...810...25G} constructed three-dimensional extinction maps of the Milky Way using stellar photometry and colors, providing valuable insights into the spatial distribution of dust. With the release of Gaia data \citep{2016A&A...595A...2G}, the precise parallaxes and proper motions significantly improved the accuracy of stellar distance measurements, leading to the development of many more high-resolution 3D extinction maps \citep{2019ApJ...887...93G,2021ApJ...906...47G,2022arXiv220411715L}. These extinction maps are primarily derived from IR and optical observations. \citet{2021ApJS..254...38S} (hereafter S21) expanded on \citet{2018ApJ...861..153S} by improving the blue-edge method \citep{2001ApJ...558..309D,2017AJ....153....5J} and using data from GALEX, GAIA, LAMOST, and GALAH to derive NUV color excesses for over 1 million sources, resulting in a NUV extinction map that covers approximately one-third of the sky. However, the limited data volume from spectral surveys such as LAMOST \citep{Luo2015}, GALAH \citep{Buder2021}, and APOGEE \citep{SDSSDR17} significantly restricts the availability of ultraviolet color excess information. \citet{2023MNRAS.524.1855Z} (hereafter Z23) develop, validate, and apply a forward model to estimate stellar atmospheric parameters for 220 million stars with XP spectra from Gaia DR3, representing an increase of more than an order of magnitude compared to traditional spectroscopic surveys. The stellar parameters provided by Z23 have the potential to greatly increase the number of UV color excess measurements, expanding from the million-source scale to the tens of millions.

\cite{1989ApJ...345..245C}, \cite{1994ApJ...422..158O}, and \cite{1999PASP..111...63F} all report extinction curves based on \rv=\av/\ebv. \cite{2019ApJ...877..116W} determined the optical-to-mid-infrared extinction ${A_\lambda}$$/$${A_{G_{\rm RP}}}$ for 21 wavebands. The extinction model obtained from optical to mid-infrared is consistent with the observed extinction law to within 2.5\%. In UV bands, \cite{2013ApJ...771...68P} measured the extinction curve at high latitudes and found that there is no unique \rv\ that can simultaneously satisfy the extinction in the optical and UV bands at high latitudes. \cite{2018ApJ...861..153S} derive NUV color excesses for over 25,000 sources and FUV color excesses for more than 4,000 sources using data from GALEX, APASS, and LAMOST. The resulting extinction law is consistent with \citet{1999PASP..111...63F} (F99) for \rv = 3.35. \cite{2023MNRAS.525.2701Y} analyze UV extinction coefficients for over 20,000 common sources from UVOT, 2MASS, and LAMOST by measuring \uvwa, \uvmb, and \uvwb\ color excesses, finding consistency with F99 for \rv = 3.0. \citet{2021ApJS..256...46S} found through their study of UV color excess ratios in high Galactic latitude molecular clouds that low-extinction molecular clouds may have a higher proportion of smaller dust grains, a finding later confirmed by \citet{2024arXiv240806986S}. Thanks to current large-scale surveys, constructing an \rv\ map has become possible. \citet{2016ApJ...821...78S} create a \rv\ map across the nearby Galactic disk (within 4 kpc) using APOGEE spectroscopic data and multi-band photometry from the optical to the near-infrared. \citet{2023ApJS..269....6Z} (hereafter ZYC23) construct a two-dimensional \rv\ map of the Milky Way based on optical-to-mid-infrared color excess measurements developed by \citet{2023ApJS..264...14Z} (hereafter ZY23). More recently, \citet{2025Sci...387.1209Z} (hereafter ZG25) achieve high-precision three-dimensional mapping of extinction curve variations across the Milky Way and the Magellanic Clouds using low-resolution Gaia XP spectra for the first time. However, their measurements did not include UV color excess information. The UV color excess data obtained in this study will provide strong constraints on extinction laws. 

This article is structured as follows. Section~\ref{data} describes the data that will be used. Section~\ref{intr} describes how to establish the relationship between intrinsic colors and stellar parameters. Section~\ref{color} acquires UV color excess information on the scale of tens of millions of stars and creates UV extinction maps. Section~\ref{ratio} discusses the color excess ratios and matches with the extinction curves. Finally, the last section concludes the paper, summarizing the key results.

\section{Data} \label{data}

This study primarily utilizes stellar parameters (\teff, \feh, and \logg) provided by spectral surveys, combined with UV photometric data from GALEX GR6/7 and SWIFT/UVOT, as well as optical photometric data from Gaia. The aim is to establish the relationship between stellar parameters and intrinsic ultraviolet color indices using the blue-edge method and to obtain ultraviolet color excess data for the relevant bands of GALEX and UVOT. The data used in this study are as follows.

\subsection{Photometric Data from the Gaia, GALEX and UVOT} \label{subsec:tables}

Gaia Data Release 3 (Gaia DR3) includes astrometric measurements of proper motion and parallax, as well as photometric measurements in three optical bands: the white-light G band (330--1050 nm), the blue (\bp), and the red (\rp) bands. These measurements cover 1.8 billion sources with a nominal survey limit of G = 20.7 mag. \citep{2016A&A...595A...2G,2022yCat.1355....0G}. This work primarily utilizes the photometric data from the Gaia DR3 \bp\ and \rp\ bands.

GALEX GR/6+7 \citep{2014Ap&SS.354..103B} provides the largest high-quality photometric database in UV bands. It is equipped with a 50 cm primary mirror and a focal length of 299.8 cm. The two filters of GALEX are FUV ($\lambda_{\rm eff} \sim 1528\ \AA$) and NUV ($\lambda_{\rm eff} \sim 2310\ \AA$). The catalog we use is GUVcat\_AIS\_fov055, with the total number of unique AIS sources (eliminating duplicate measurements) being 82,992,086, and a typical depth of FUV = 19.9 and NUV = 20.8 AB mag \citep{2017ApJS..230...24B}. All the GALEX data used in this paper can be found in MAST (\citealt{GUVcat_GUVmatch}, \dataset[10.17909/t9-pyxy-kg53]{http://dx.doi.org/10.17909/t9-pyxy-kg53}). 

The UV/Optical Telescope (UVOT; \citealt{2005SSRv..120...95R}) is one of three instruments flying aboard the Swift spacecraft \citep{2004ApJ...611.1005G}. The UVOT is designed to capture early (approximately 1 minute) ultraviolet and optical photons from the afterglow of gamma-ray bursts in the 170--600 nm band and to perform long-term observations of these afterglows. It features a 30 cm primary mirror and a 7.2 cm secondary mirror, with a primary f-ratio of f/2.0 that increases to f/12.72 after the secondary mirror. The total number of sources in the UVOT catalog is 13,860,568 \citep{2015yCat.2339....0Y}. It provides photometric measurements in six bands: $v$, $b$, $u$, $uvw1$, $uvm2$, $uvw2$, with central wavelengths of 546.8 nm, 439.2 nm, 346.5 nm, 260.0 nm, 224.6 nm, and 192.8 nm, respectively. In this work, we use the three UV bands: $uvw1$, $uvm2$, $uvw2$.

\subsection{Stellar parameters from spectroscopic surveys}

The stellar parameters (\teff, \feh, and \logg) used in this study are derived from LAMOST, GALAH, APOGEE, and Z23. LAMOST \citep{Luo2015} surveys the entire available northern sky and can capture up to 4000 spectra in a single exposure, achieving a limiting magnitude of 19 mag in the r band with a resolution of R = 1800 \citep{2012RAA....12..723Z}. The DR11 catalog (L11) includes a total of 7,774,147 stars. The GALAH survey \citep{Buder2021} primarily focuses on the southern hemisphere and employs the HERMES spectrograph to simultaneously collect high-resolution spectra (R $\sim$ 28,000) for 360 stars. The DR3 catalog (G3) contains 588,571 unique stars. The APOGEE survey \citep{SDSSDR17} utilizes a fiber spectrograph that records 300 spectra at a spectral resolution of R $\sim$ 22,500. The DR17 catalog (A17) includes a total of 733,901 stars.

Z23 obtained precise atmospheric parameter measurements for over 220 million stars using Gaia XP spectra data. They trained their model using data from the LAMOST survey and incorporated near-infrared photometry from 2MASS and WISE. By constraining the model at longer wavelengths, they were able to distinguish between stellar temperature and extinction. This method fully leveraged all available data, significantly reducing degeneracies among stellar parameters and overcoming many of the major limitations faced by the astrophysical parameters released in Gaia DR3. As a result, they estimated the atmospheric parameters (\teff, \feh, and \logg), corrected distances, and extinction for more than 220 million stars with Gaia DR3 XP spectra. The typical errors in their study are 90 K for \teff\ and 0.15 dex for \feh\ and \logg. This study primarily uses the atmospheric parameters (\teff, \feh, and \logg) provided by Z23.

\subsection{Catalog combination and source selection}
The cross-matching process involves several steps. First, the stellar parameters (\teff, \feh, and \logg) obtained from spectroscopic surveys are matched with Gaia DR3 to acquire data in the Gaia/\bp\ and Gaia/\rp\ bands, along with distance information. Subsequently, these results are cross-matched with GALEX GR6/7 and SWIFT/UVOT within a $3^{\prime\prime}$ radius to obtain UV band information. This process generates eight combined catalogs: L11-Gaia-GALEX (``LGG'' hereafter, 2,274,770 stars), G3-Gaia-GALEX (``GGG'' hereafter, 279,590 stars), A17-Gaia-GALEX (``AGG'' hereafter, 208,972 stars), Z23-Gaia-GALEX (``ZGG'' hereafter, 24,201,863 star) and L11-Gaia-UVOT (``LGU'' hereafter, 68,513 stars), G3-Gaia-UVOT (``GGU'' hereafter, 7,251 stars), A17-Gaia-UVOT (``AGU'' hereafter, 10,288 stars), Z23-Gaia-UVOT (``ZGU'' hereafter, 2,083,484 stars). These directories are used to study color excesses in the UV band.

To ensure the quality of the selected sources, strict quality criteria are applied to the cross-matched catalog. High-confidence sources are chosen from Z23, requiring $teff\_confidence$, $feh\_confidence$, $logg\_confidence$ values greater than 0.5 and quality\_flags $<$ 8. The relative uncertainty in \teff\ is required to be less than 5\% ($\sigma_{T_{\rm eff}}/T_{\rm eff}<5\%$). The range of \feh\ is limited to between $-2\ \text{dex}$ and $0.5\ \text{dex}$, with an error margin not exceeding 0.3 dex. Furthermore, the accuracy of both \bp\ and \rp\ magnitudes is required to be better than 0.02 mag.

For the GALEX-related combined catalogs, we limit the accuracy of NUV and FUV band to be better than 0.30 mag. Additionally, the brightness is limited to ${\rm NUV} > 13.85$ mag and ${\rm FUV} > 13.73$ mag to avoid issues caused by UV-bright sources \citep{2017ApJS..230...24B}. The \teff\ is limited to $>4500\ {\rm K}$ for the NUV band and $>7000\ {\rm K}$ for the FUV band. For the UVOT-related combined catalog, the accuracy of \uvwa, \uvmb, and \uvwb\ are all controlled at better than 0.30 mag. The \teff\ is limited to $>4500\ {\rm K}$. The final selected combined catalogs have 93,230,392 sources in \bp\rp\ band, 11,915,943 sources in NUV band (1,842,413 in LGG, 236,337 in GGG, 158,574 in AGG and 9,678,619 in ZGG), 87,782 sources in FUV band (39,698 in LGG, 2,507 in GGG and 45,577 in ZGG)，379,338 sources in \uvwa\ band (29,176 in LGU, 2,771 in GGU, 3,265 in AGU and 344,126 in ZGU), 162,206 sources in \uvmb\ band (19,275 in LGU, 2,069 in GGU, 1,979 in AGU and 138,883 in ZGU), and 293,172 sources in \uvwb\ band (28,796 in LGU, 2,923 in GGU, 3,513 in AGU and 257,940 in ZGU). 

\section{Determination of the intrinsic color index}  \label{intr}
Color excess can be determined by subtracting the intrinsic color index from the observed color index, making the accurate determination of intrinsic color indices crucial for calculating color excess. In this section, we will use the ``blue edge'' method \citep{2001ApJ...558..309D,2017AJ....153....5J} to identify zero/low extinction sources. The selected zero/low extinction sample will then be used as a training set for a random forest regression model to establish the relationship between intrinsic color indices and stellar parameters. To provide a clear explanation of the methodology used in this work, we will use the Z23-related catalog, which contains the largest number of sources, as a representative for the following descriptions and discussions. 

\subsection{Determination of the representative intrinsic color index in the UV bands}
The ``blue edge'' method suggests that for stars with the same spectral type, zero/low extinction stars can be identified by finding the bluest stars in terms of color index, and their intrinsic color indices can be obtained accordingly. For the UV bands, the intrinsic color index of stars is not only dependent on \teff\ but also influenced by \feh\ \citep{2018ApJ...861..153S, 2021ApJS..254...38S}. Moreover, compared to previous studies, this work also includes giant stars in the study. Therefore, it is necessary to consider the effects of \teff, \feh, and \logg\ on the intrinsic color indices in the UV bands. Additionally, chromospheric activity and higher uncertainties in the UV bands might affect the determination of intrinsic color indices. As a result, we initially apply the "blue edge" method to the optical bands, which provide more accurate photometry and are less affected by chromospheric activity, to identify zero/low extinction sources.

\begin{figure*}
\centering
\includegraphics[width=1\textwidth,angle=0]{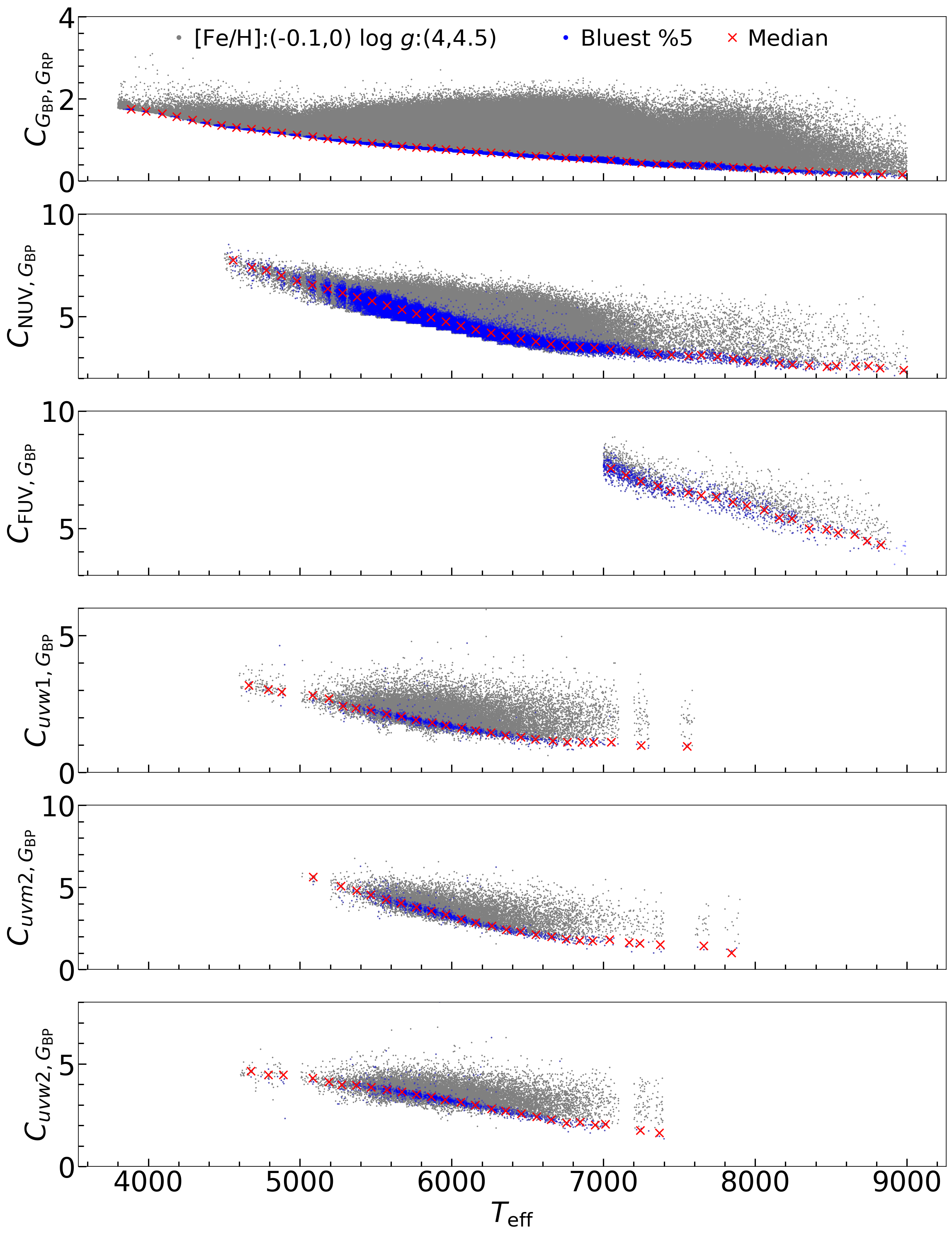}
\caption{The change of the color index with \teff for the stars with $-0.1<$\feh$<0$ and $4<$\logg$<4.5$. The grey dots denote all the stars, the blue dots denote the selected zero/low extinction sources, and the red crosses represent the median of zero/low extinction sources in each \teff\ bin of 100 K.}
\label{lby}
\end{figure*}

Specifically, we divide the every final selected combined catalogs into 100 K, 0.1 dex and 0.5 dex intervals based on \teff, \feh\ and \logg. For each stellar parameter interval, the bluest 5\% of stars, after performing $3\sigma$ clipping in the optical band within the corresponding interval, are considered zero or low-extinction sources. The median intrinsic color index of these zero or low-extinction stars is then used as the representative value for that parameter interval. As mentioned above, the zero/low extinction sources for the UV bands are selected based on the optical band, and the median UV color index of these zero/low extinction sources in each parameter interval is taken as the representative value for the intrinsic UV color index. If the median in intervals with few sources becomes unreliable, it can be adjusted by replacing intervals that deviate more than 1 $\sigma$ from their neighbors with the mean of the surrounding 3 $\times$ 3 $\times$ 3 block. Figure~\ref{lby} illustrates the variation of color indices within the range of $-0.1<$\feh$<0$ and $4<$\logg$<4.5$ with \teff. In general, zero/low extinction sources selected in the optical band show a good fit to the blue edge in UV band. 

\subsection{The Relationship between Intrinsic Color Index and Stellar Parameters}
A machine learning algorithm called Random Forest regression \citep{2011JMLR...12.2825P} is employed to construct a model that relates intrinsic color indices to stellar parameters. Random Forest regression, a meta-estimator, fits multiple decision trees to subsets of the dataset and uses averaging to enhance prediction accuracy while controlling overfitting. A grid of representative intrinsic color index values selected from zero-extinction sources serves as the input data. For training the Random Forest model, representative values are used across a total of 6,919, 5,916, 2,031, 3,457, 3,109, and 3,418 parameter intervals corresponding to the optical, NUV, FUV, \uvwa, \uvmb, and \uvwb\ bands for Z23-related catalogs, respectively. 

\begin{figure*}
\centering
\includegraphics[width=1\textwidth,angle=0]{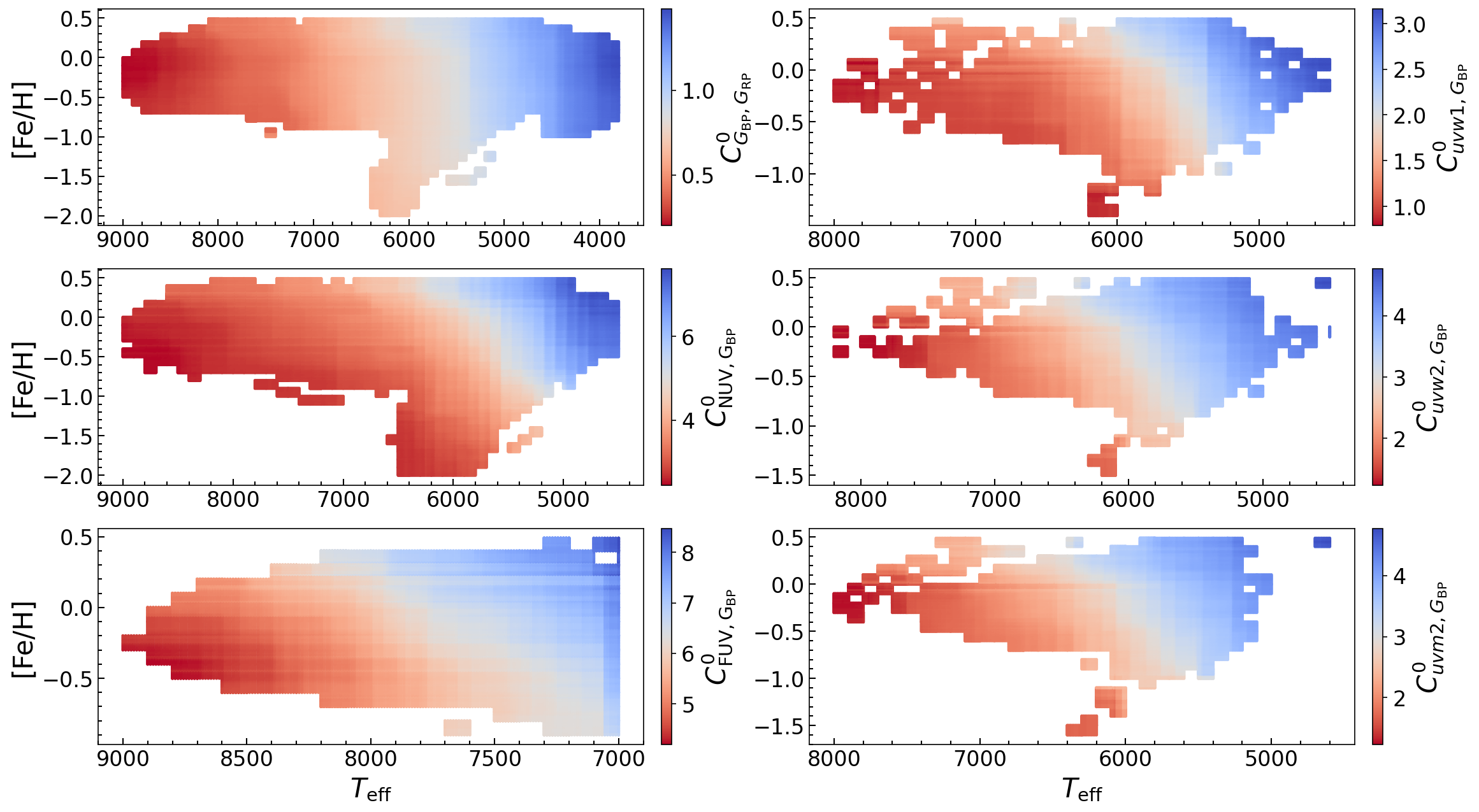}
\caption{The variation of the intrinsic color indices $C_{G_{\rm BP},G_{\rm RP}}^0$ (top-left), $C_{{\rm NUV},G_{\rm BP}}^0$ (middle-left), $C_{{\rm FUV},G_{\rm BP}}^0$ (bottom-left), $C_{uvw1,G_{\rm BP}}^0$ (top-right), $C_{wum2,G_{\rm BP}}^0$ (middle-right), and $C_{uvw2,G_{\rm BP}}^0$ (bottom-right) with \teff\ and \feh\ at \logg\ = 4 dex.}
\label{intri2}
\end{figure*}

\begin{figure*}
\centering
\includegraphics[width=1\textwidth,angle=0]{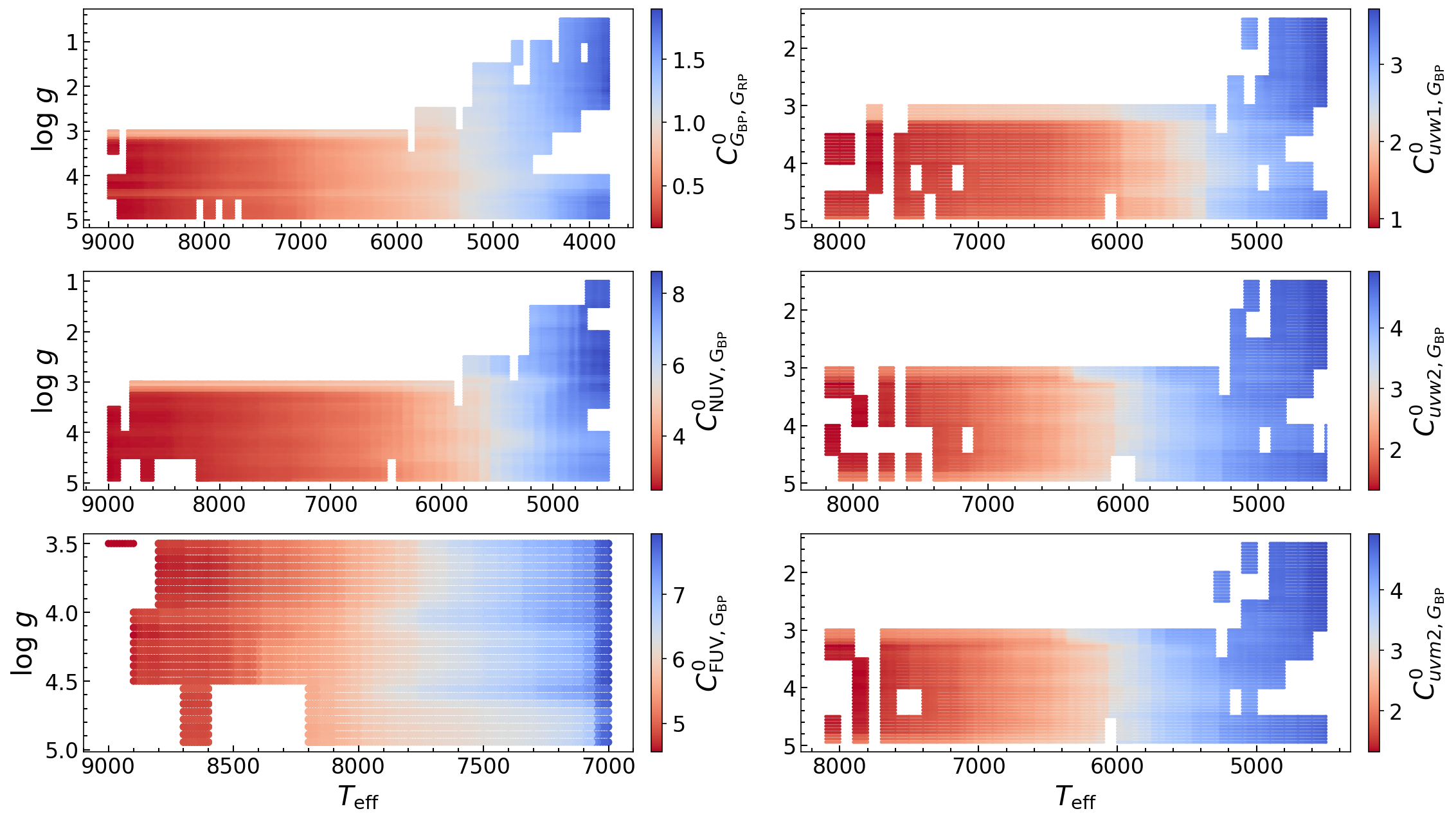}
\caption{The same as Fig.~\ref{intri2}, but for the variation of the intrinsic color indices with \teff\ and \logg\ at \feh\ = 0 dex.}
\label{intri}
\end{figure*}

Figures~\ref{intri2} and \ref{intri} illustrate the variation of intrinsic color indices with \teff\ and \feh\ at \logg\ = 4, as well as with \teff\ and \logg\ at \feh\ = 0. It is evident that the influence of \feh\ on the intrinsic UV color index is significantly more pronounced than in the optical band, with all UV bands showing an increase in the intrinsic color index as \feh\ increases. Additionally, \logg\ also affects the intrinsic UV color index, with giants exhibiting larger intrinsic color indices compared to dwarfs.
\subsection{Comparison of intrinsic color indices with PARSEC}
\begin{figure*}
\centering
\includegraphics[width=1\textwidth,angle=0]{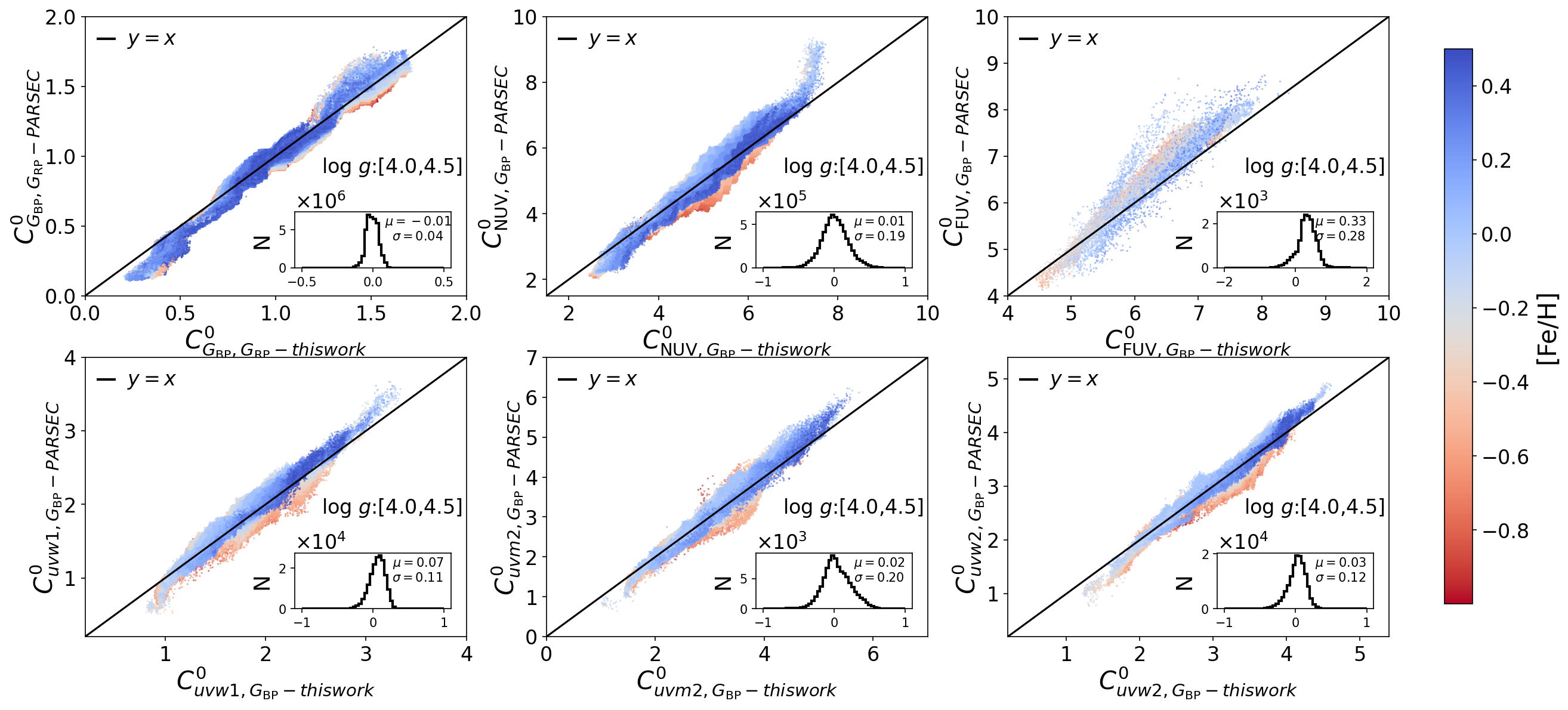}
\caption{The comparison of the intrinsic colors of dwarf stars (4$<$\logg$<$4.5) derived from this work (x-axis) and the PARSEC stellar model (y-axis). The color of the dots represents \feh, and the black line indicates the line of equality between x and y. The inset shows the distribution of the differences. }
\label{pars}
\end{figure*}

\begin{figure*}
\centering
\includegraphics[width=1\textwidth,angle=0]{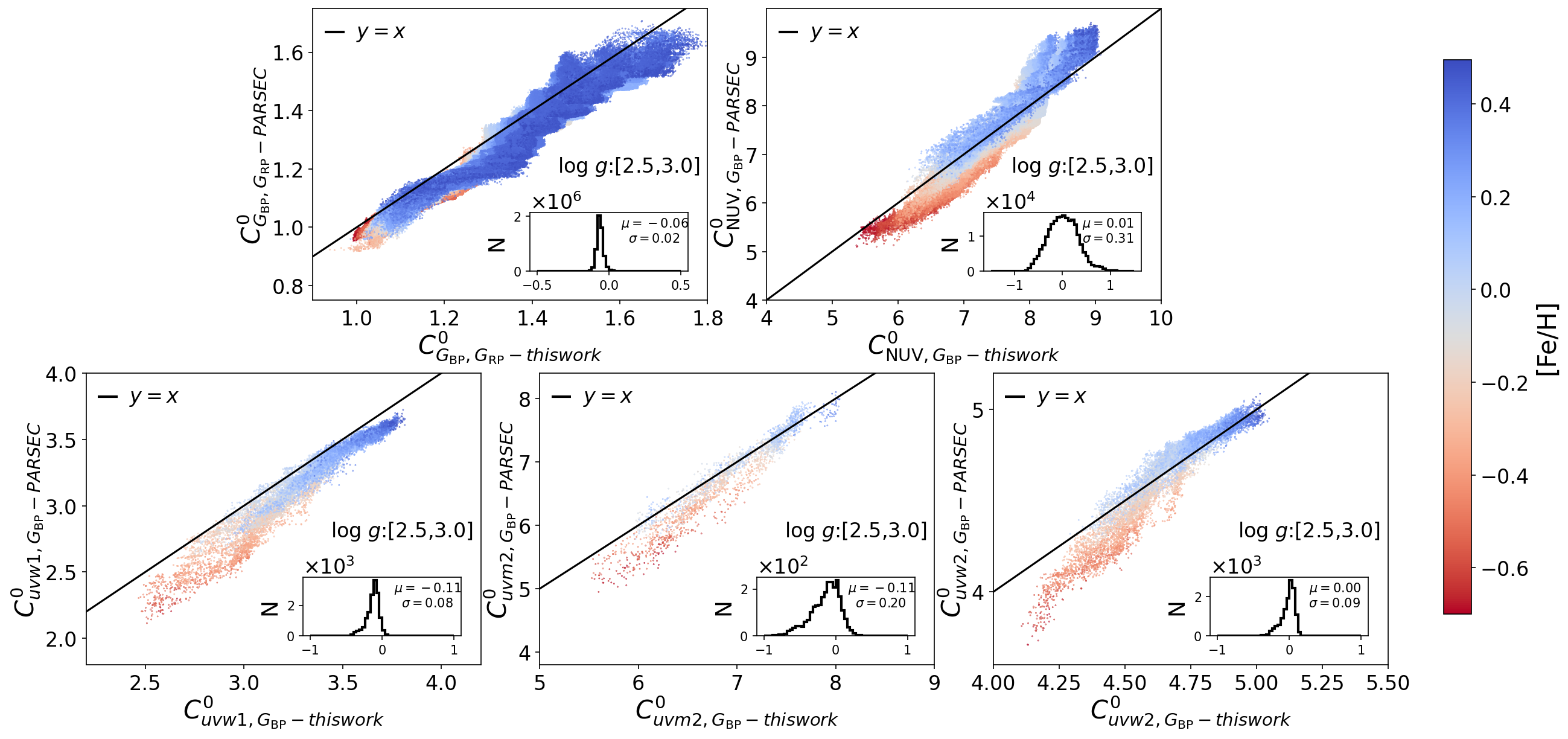}
\caption{The same as Fig.~\ref{pars}, but for giant stars (2.5$<$\logg$<$3).}
\label{pars2}
\end{figure*}

Figures~\ref{pars} and~\ref{pars2} compare the intrinsic color indices of typical dwarf stars (4 $<$ \logg\ $<$ 4.5) and giant stars (2.5 $<$ \logg\ $<$ 3) obtained in this work with those from the PAdova and TRieste Stellar Evolution Code (PARSEC; \citealt{2012MNRAS.427..127B}). For the optical bands, the intrinsic color indices for both dwarfs and giants in this work show good overall agreement with the PARSEC results, with dispersions of 0.04 mag for dwarfs and 0.02 mag for giants. There is no obvious trend in the differences as a function of \feh. In the UV bands, except for the FUV band, the intrinsic color indices of dwarfs also show good agreement with PARSEC, with dispersions of 0.19 mag, 0.28 mag, 0.11 mag, 0.20 mag, and 0.12 mag for the NUV, FUV, \uvwa, \uvmb, and \uvwb\ bands, respectively. While the median differences are close to zero for most bands, the FUV band shows a significantly larger median difference of 0.33 mag. Additionally, in all UV bands, the intrinsic color indices provided by PARSEC are noticeably redder at larger $C^0$ values (for relatively cooler stars). This trend is particularly pronounced in the GALEX-related bands. The differences also show no clear trend with \feh, except in the FUV band, where for larger $C^0$ values, metal-rich stars exhibit significantly larger discrepancies. For giant stars in the UV bands, there are no FUV results due to the \teff\ $> 7000$ K constraint. The dispersions of the differences for the NUV, \uvwa, \uvmb, and \uvwb\ bands are 0.31 mag, 0.08 mag, 0.20 mag, and 0.09 mag, respectively, with small median differences similar to the results for dwarfs. In the NUV band, PARSEC shows a slightly redder trend for metal-rich sources. However, in the UVOT-related bands, PARSEC displays a slightly bluer trend for sources with lower \feh.

\section{The Color Excess} \label{color}

\subsection{Determination of Ultraviolet Color Excess}
The color excess is calculated by subtracting the intrinsic color index, derived from stellar parameters, from the observed color index of the stars. Leveraging the extensive stellar parameters provided by LAMOST, GALAH, APOGEE and Z23, this work has significantly increased the quantity of UV color excess data compared to previous studies. Specifically, it has obtained $E_{G_{\rm BP},G_{\rm RP}}$ for 92,142,820 sources, $E_{{\rm NUV},G_{\rm BP}}$ for 11,624,802 sources, and $E_{{\rm FUV},G_{\rm BP}}$ for 65,531 sources. In comparison, S21 provided $E_{G_{\rm BP},G_{\rm RP}}$ for 4,169,769 sources, $E_{{\rm NUV},G_{\rm BP}}$ for 1,244,504 sources, and $E_{{\rm FUV},G_{\rm BP}}$ for 56,123 sources, with $E_{G_{\rm BP},G_{\rm RP}}$ increasing by over 20 times and $E_{{\rm NUV},G_{\rm BP}}$ nearly 10 times,  showing a substantial improvement in the number of sources analyzed. For UVOT-related bands, this work has obtained $E_{{\rm uvw1},G_{\rm BP}}$ for 336,633 sources, $E_{{uvm2},G_{\rm BP}}$ for 137,739 sources, and $E_{{uvw2},G_{\rm BP}}$ for 253,271 sources. This study offers several key advancements over S21: (1) Expanded the \teff\ range for obtaining intrinsic UV color indices, (2) the consideration of giant stars (\logg $<$3.5), (3) The stellar parameters are sourced from spectroscopic surveys like LAMOST, GALAH, and APOGEE, as well as from Z23 based on Gaia XP spectra, which significantly increases the number of UV sources with available stellar parameters, and (4) the acquisition of UV color excess data for UVOT-related bands.

\subsection{Color-Excess Errors}  

\begin{figure*}
\centering
\includegraphics[width=1\textwidth,angle=0]{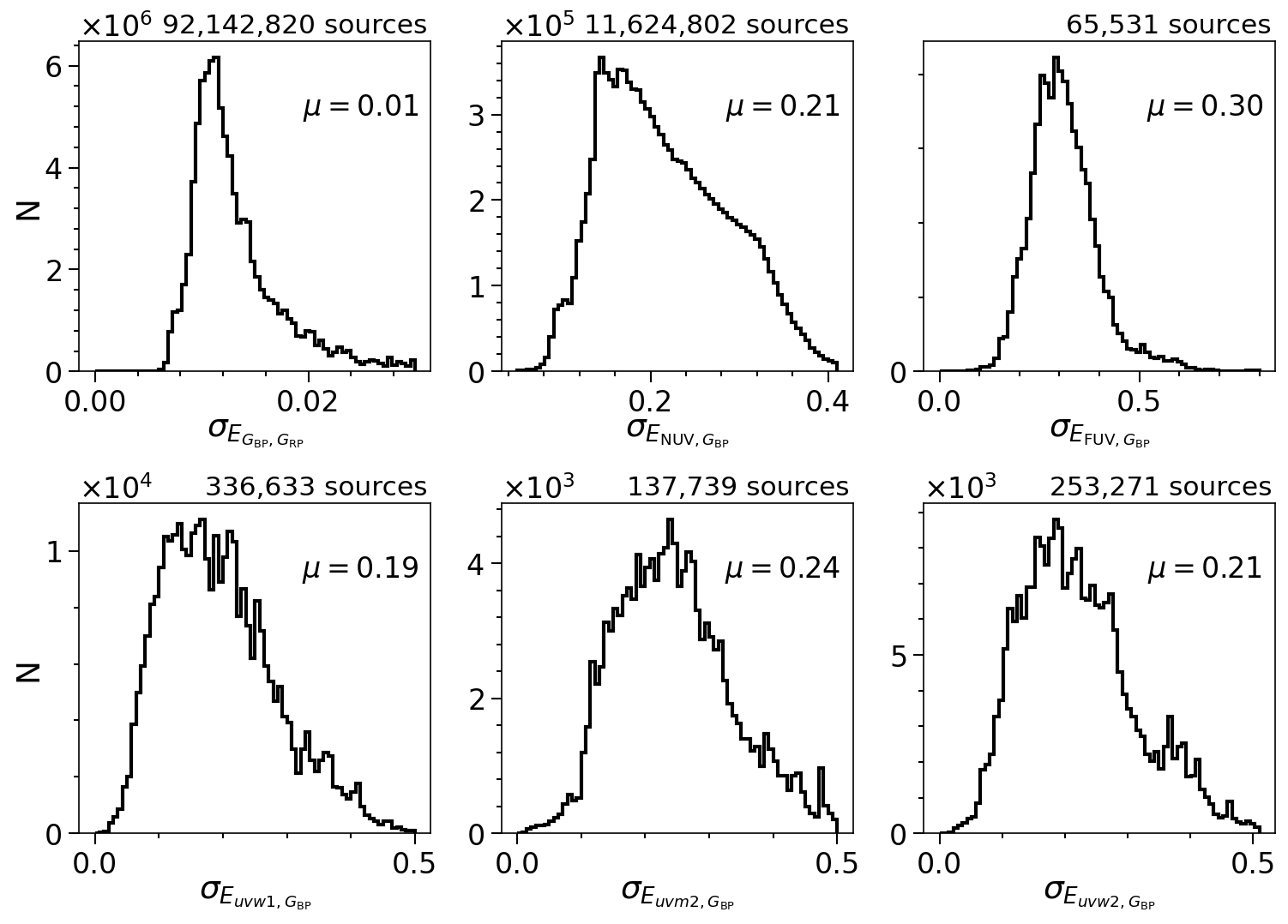}
\caption{The distribution of color excess errors, as well as the number of color excess values and the median color excess error, are labeled in each panel.}
\label{ero1}
\end{figure*}

The errors in color excesses can be estimated by combining the errors in the observed color indices and the intrinsic color indices. The error in the observed color index can be directly estimated from the photometric error in each band, while the error in the intrinsic color index can be estimated from the dispersion of zero/low extinction sources within the corresponding parameter range. In fact, using the random forest method to estimate the intrinsic color index takes into account the variation within the parameter range, so the error in the intrinsic color index is smaller than the dispersion of zero or low-extinction sources S21. Figure~\ref{ero1} shows the error distribution of the color excess in the optical and UV bands. The median errors in color excess are 0.01 mag for $E_{G_{\rm BP},G_{\rm RP}}$, 0.21 mag for $E_{{\rm NUV},G_{\rm BP}}$,  0.30 mag for $E_{{\rm FUV},G_{\rm BP}}$, 0.19 mag for $E_{{\rm uvw1},G_{\rm BP}}$, 0.24 mag for $E_{{\rm uvm2},G_{\rm BP}}$, 0.21 mag for $E_{{\rm uvw2},G_{\rm BP}}$.
 
\begin{figure*}
\centering
\includegraphics[width=1\textwidth,angle=0]{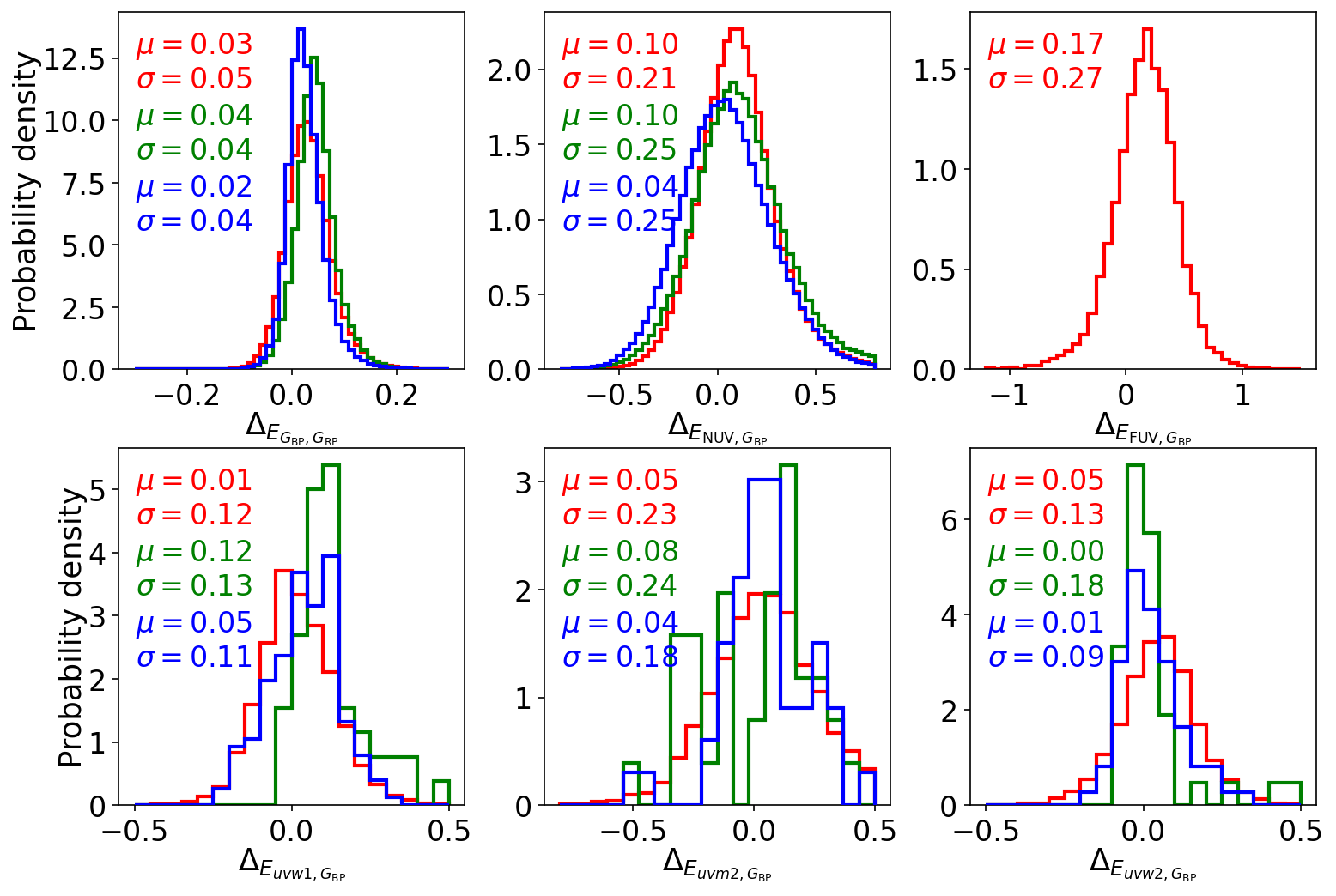}
\caption{The distribution of color excess differences between Z23 and common sources from LAMOST (red), GALAH (green), and APOGEE (blue) respectively. The median and dispersion of the color excess differences, are labeled in each panel.}
\label{ero2}
\end{figure*}

The dispersion in the color excess differences between Z23 related combined catalogs and L11, G3, A17 related combined catalogs can also be used to estimate the error in the color excess as shown in Figure~\ref{ero2}. The dispersion values are 0.04 mag - 0.05 mag for $E_{G_{\rm BP},G_{\rm RP}}$, 0.21 mag - 0.25 mag for $E_{{\rm NUV},G_{\rm BP}}$, 0.27 mag for $E_{{\rm FUV},G_{\rm BP}}$, 0.11 mag - 0.13 mag for $E_{{\rm uvw1},G_{\rm BP}}$, 0.18 mag - 0.24 mag for $E_{{\rm uvm2},G_{\rm BP}}$, and 0.09 mag - 0.18 mag for $E_{{\rm uvw2},G_{\rm BP}}$, which are in good agreement with the estimated errors.

\subsection{Comparison with Previous Results}
\begin{figure*}
\centering
\includegraphics[width=1\textwidth,angle=0]{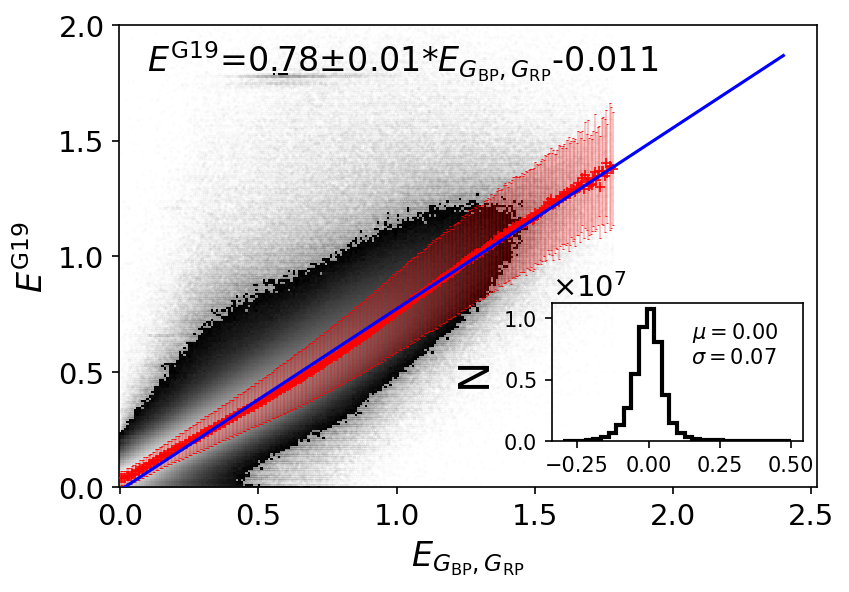}
\caption{The comparison of $E_{G_{\rm BP},G_{\rm RP}}$ derived in this work with $E^{\rm G19}$ (the reddening) from G19. The red plus signs and error bars represent the median and dispersion of color excess for each 0.005 mag bin of \ebr. The inset shows the distribution of the differences.}
\label{sydb}
\end{figure*}

Figure~\ref{sydb} shows the comparison of the optical band color excess (\ebr) from this work with the reddening ($E^{\rm G19}$) from \citet{2019ApJ...887...93G} (G19). It is evident that \ebr\ from this work is in good agreement with $E^{\rm G19}$, with the median difference being 0.00 and the scatter being 0.07 mag. The ratio $E^{\rm G19} / E_{G_{\rm BP},G_{\rm RP}}$ obtained in this work is 0.78, which is slightly larger than the value of 0.66 reported by S21.

\begin{figure*}
\centering
\includegraphics[width=1\textwidth,angle=0]{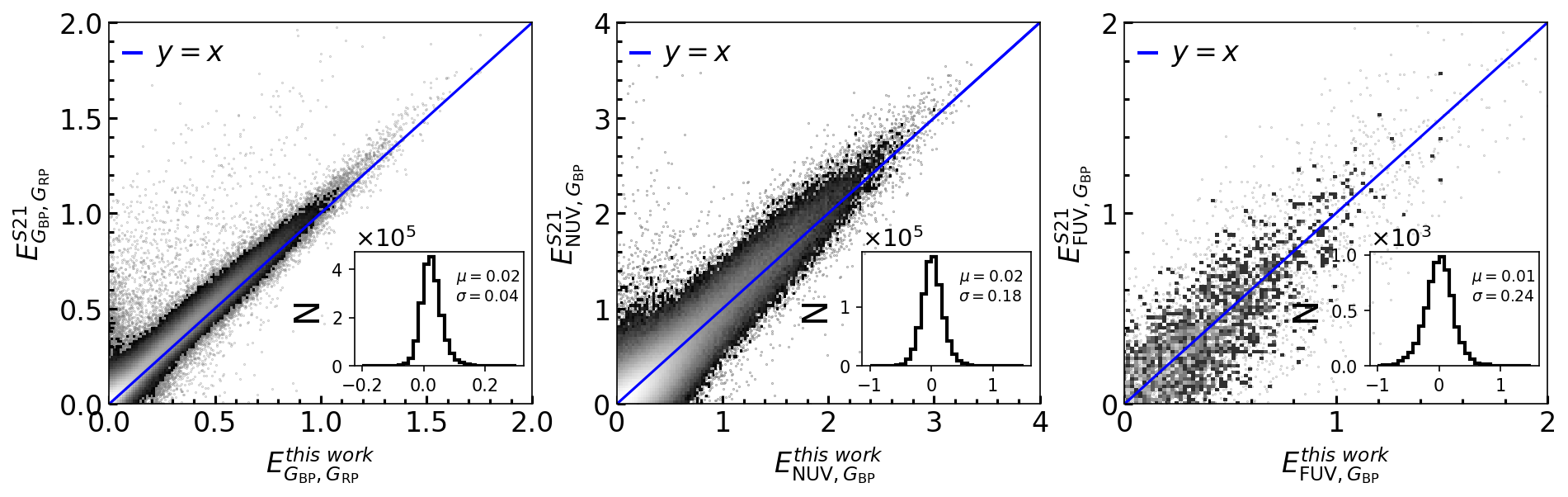}
\caption{The comparison of $E_{G_{BP},G_{RP}}$, $E_{{\rm NUV},G_{BP}}$ and $E_{{\rm FUV},G_{BP}}$ derived in Z23 with those from S21. The blue line indicates the line of equality between x and y and the inset shows the distribution of the differences.}
\label{ds}
\end{figure*}

Figure~\ref{ds} presents a comparison of the color excesses in the optical, NUV, and FUV bands from this work with those from S21. The results demonstrate strong consistency between the two datasets. The median differences for the optical, NUV, and FUV bands are 0.02, 0.02, and 0.01, respectively, all of which are very close to 0. The scatter in the differences for these bands is 0.04, 0.18, and 0.24, respectively, which is comparable to the errors in the color excess measurements.

\subsection{The Ultraviolet Extinction Map}
\begin{figure*}
\centering
\includegraphics[width=1\textwidth,angle=0]{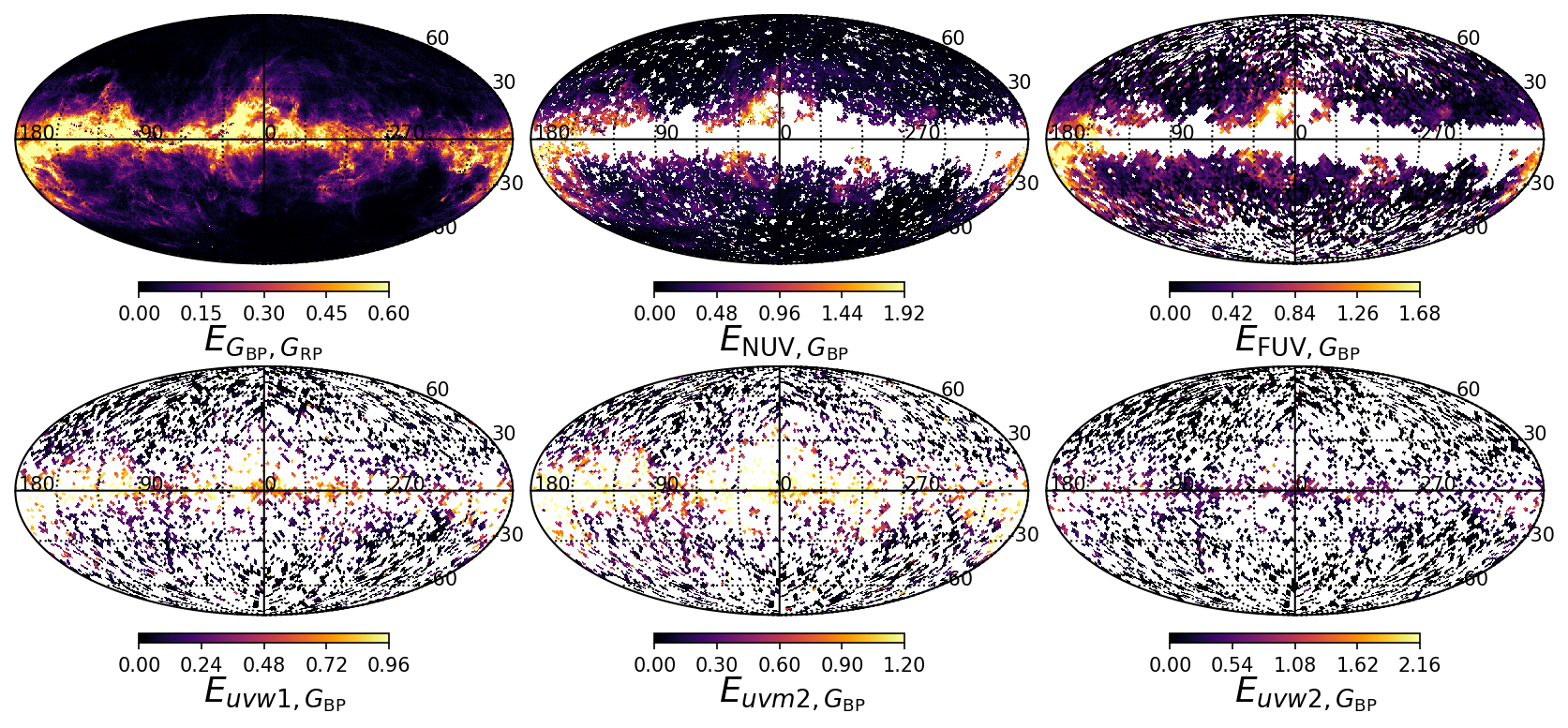}
\caption{Gridding map by the HEALPix method of $E_{G_{\rm BP},G_{\rm RP}}$ (top-left), $E_{{\rm NUV},G_{BP}}$ (top-middle), $E_{{\rm FUV},G_{BP}}$ (top-right), $E_{uvw1,G_{BP}}$ (bottom-left), $E_{uvm2,G_{BP}}$ (bottom-middle) and $E_{uvw2,G_{BP}}$ (bottom-right).}
\label{xgt1}
\end{figure*}

Using the vast amount of color excess data mentioned above, we can create separate UV extinction maps for the optical, NUV, FUV, \uvwa, \uvmb, and \uvwb\ bands. For this purpose, we use the HEALPix algorithm to pixelize the celestial sphere \citep{2005ApJ...622..759G} , and employ the "healpy" Python module \footnote{\url{https://healpy.readthedocs.io/}} to grid the data. The extinction maps for the six bands are shown in Figure~\ref{xgt1}. For the optical and NUV bands we choose $N_{side} = 128$, which divides the entire sky into 196,608 pixels , with each pixel covering approximately 0.2 deg$^2$. For the FUV, \uvwa, \uvmb, and \uvwb\ bands, we choose $N_{side} = 32$, which divides the sky into 12,288 pixels ($N_{pix}$), with each pixel covering approximately 0.84 deg$^2$. We perform an iterative 3$\sigma$ clipping and take the median value in each bin as the representative value. It can be seen that the optical band extinction map almost covers the entire sky, clearly displaying the Galactic disk and various molecular cloud features. The large-scale structure of the UV extinction map aligns well with that of the optical bands. The NUV and FUV extinction maps effectively showcase the extinction features at high galactic latitudes and some small-scale structures, as well as numerous molecular cloud structures in both the galactic center and anti-center directions. In the \uvwa, \uvmb, and \uvwb extinction maps, there is a noticeable increase in extinction towards the galactic plane, clearly revealing the main structural features in the direction of the galactic disk.

\begin{figure*}
\centering
\includegraphics[width=1\textwidth,angle=0]{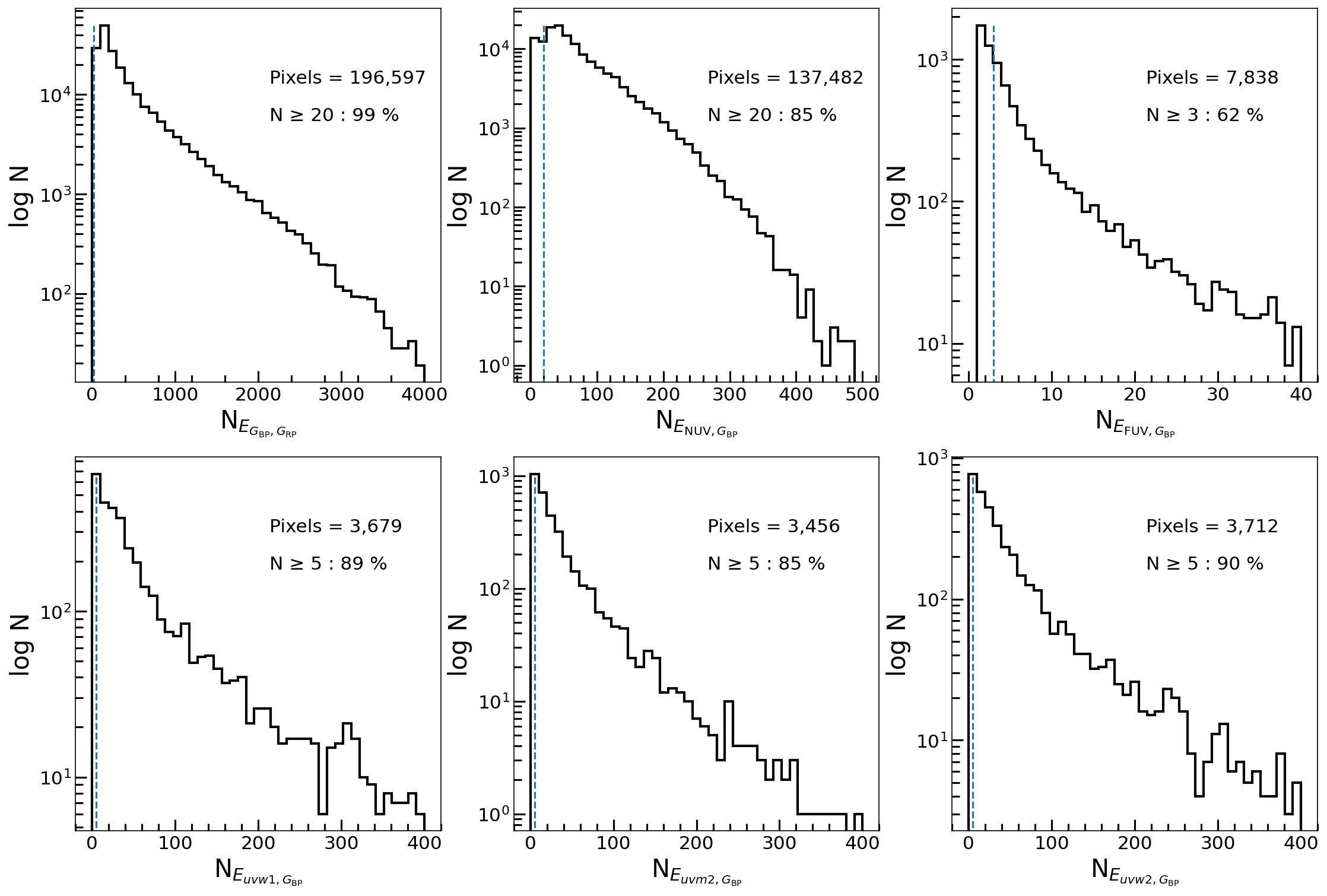}
\caption{The histogram of the number of sources in each HEALPix pixel for $E_{G_{\rm BP},G_{\rm RP}}$ (top-left), $E_{{\rm NUV},G_{BP}}$ (top-middle), $E_{{\rm FUV},G_{BP}}$ (top-right), $E_{uvw1,G_{BP}}$ (bottom-left), $E_{uvm2,G_{BP}}$ (bottom-middle) and $E_{uvw2,G_{BP}}$ (bottom-right).}
\label{zhutu1}
\end{figure*}

Figure~\ref{zhutu1} shows the histogram of the number of sources in each HEALPix pixel. For NUV extinction map, there are 137,482 pixels, covering about two-thirds of the sky, with 85\% of the pixels having more than five sources. The FUV extinction map has 7,838 pixels, covering about half of the sky, with 62\% of the pixels having more than three sources. The UV extinction maps for the \uvwa, \uvmb, and \uvwb bands have 3,679; 3,456; and 3,712 pixels, respectively, covering about one-third of the sky, with 89\%, 85\%, and 90\% of the pixels having more than five sources . In comparison, the optical band extinction map has 196,597 pixels, covering the entire sky, with 99\% of the pixels having more than five sources. 

\begin{figure*}
\centering
\includegraphics[width=1\textwidth,angle=0]{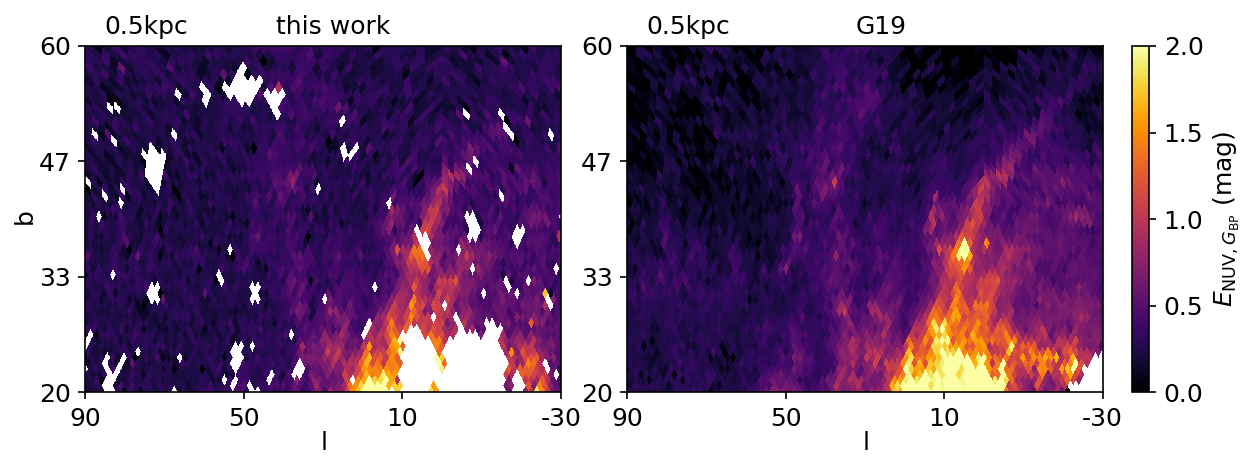}
\caption{Comparison of UV extinction maps from this work (left) and  extinction maps from G19 (right) in the region $-30^{\circ} < l < 90^{\circ}$ and $20^{\circ} < b < 60^{\circ}$ at distances $d = 0.5$ kpc.}
\label{xgtbj}
\end{figure*}

In Figure~\ref{xgtbj}, we compare the NUV extinction maps from this study with those from G19 at distances of 0.5 kpc within the region $-30^{\circ} < l < 90^{\circ}$ and $20^{\circ} < b < 60^{\circ}$. The main structure of the high-extinction region on the right side in our NUV extinction map shows excellent agreement with G19. Additionally, our NUV extinction map effectively captures the dust structures in the low-extinction region on the left. In future work, we will use the UV extinction maps from this study to further identify and discuss the dust structural features in high galactic latitude regions. 

\section{The Color Excess Ratio} \label{ratio}

\subsection{Determination of the UV color excess ratio}
\begin{figure*}
\centering
\includegraphics[width=1\textwidth,angle=0]{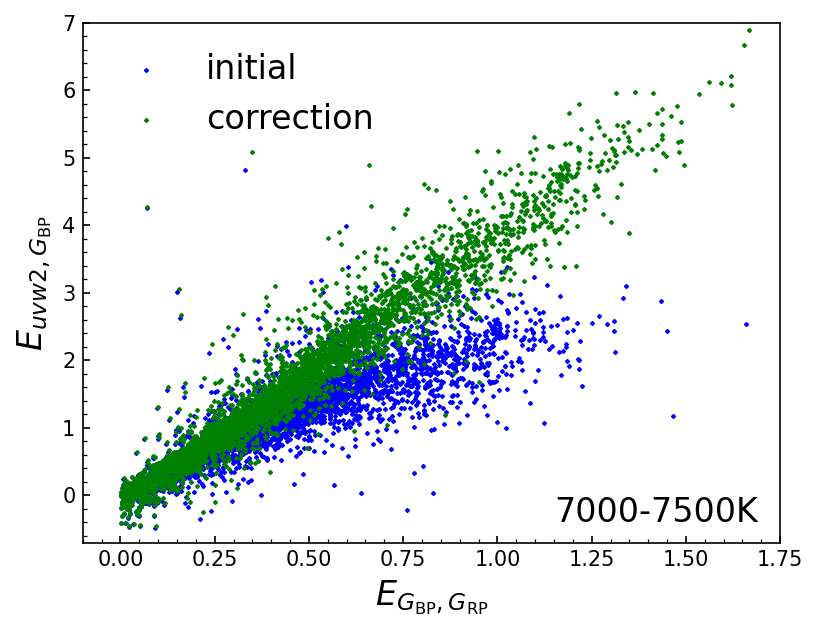}
\caption{The diagram of $E_{G_{\rm BP},G_{\rm RP}}$ vs. $E_{uvw2,G_{\rm BP}}$ for stars in the range of $7000\ {\rm K} \leq T_{\rm eff} \leq 7500\ {\rm K}$ before (blue points) and after (green points) curvature correction.}
\label{qul}
\end{figure*}

\begin{figure*}
\centering
\includegraphics[width=1\textwidth,angle=0]{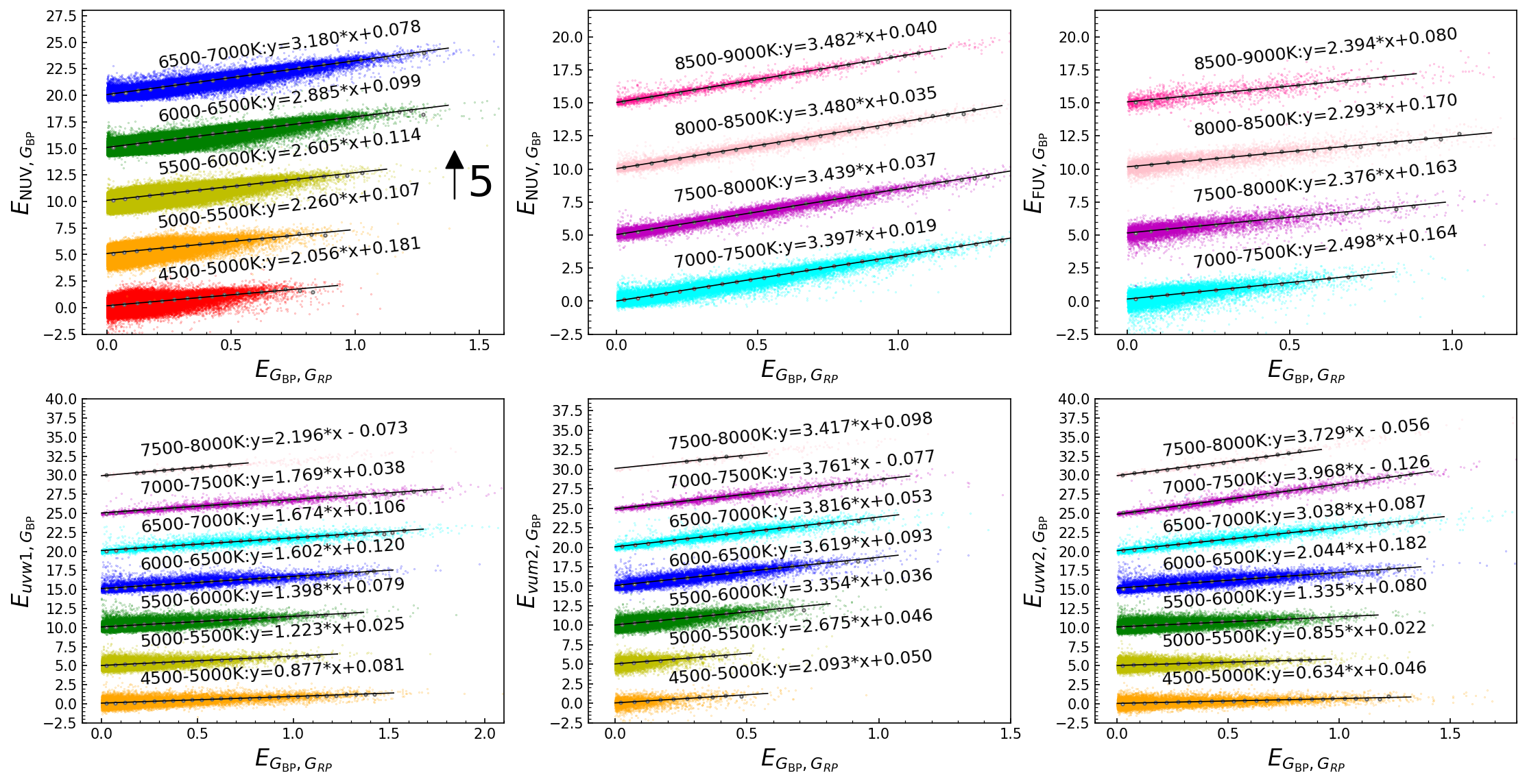}
\caption{The linear fitting of color excesses $E_{\lambda,G_{BP}}$ to $E_{G_{BP},G_{RP}}$ for different \teff\ ranges. Points in different colors represent different \teff\ ranges. The black lines indicate the best linear fit, with the fit results displayed in the upper-right corner of each panel.}
\label{sybt}
\end{figure*}

Using the color excess in the optical band and the UV bands obtained in this work, the color excess ratio (CER) can be calculated. Due to the use of broadband photometry, the CERs are affected by shifts in the effective wavelength. When performing band photometry, different spectral types (\teff) of stars and variations in extinction can lead to different degrees of effective wavelength shifts \citep{2021ApJS..254...38S,2019ApJ...877..116W,2023ApJ...956...26L}. Figure ~\ref{qul} presents an example of the CE–CE diagram for the \uvwb bands within the Teff range of 7000 to 7500 K. The blue points represent the CE–CE diagram of the original data, where the curvature of the CER is quite pronounced. This is due to the fact that when the bandwidth is not infinitely narrow, the effective wavelength of the band increases with greater extinction. The green points represent the CE–CE diagram after correcting the effective wavelength to zero extinction, showing a strong linear relationship between the ultraviolet and optical color excesses post-correction. Additionally, differences in spectral types (\teff) also contribute to variations in effective wavelength. After curvature correction, Figure~\ref{sybt} displays the fitting results for the ultraviolet CERs across different \teff\ ranges. It is evident that within the same \teff\ range, there exists a strong linear relationship between the ultraviolet and optical color excesses, although there are some differences in the CERs across different \teff\ ranges.

\begin{figure*}
\centering
\includegraphics[width=1\textwidth,angle=0]{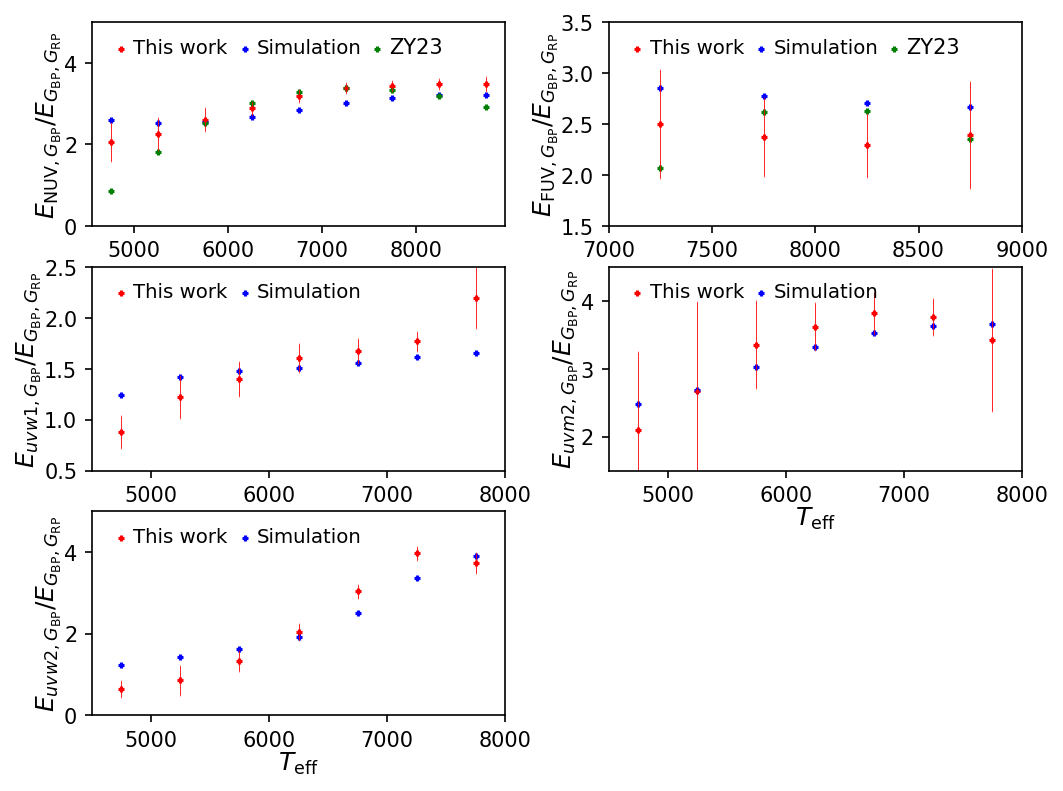}
\caption{Color excesses $E_{\lambda,G_{BP}}/E_{G_{BP},G_{RP}}$ as a function of \teff. Red points with error bars represent the color excess ratios and uncertainties from this study. Blue points indicate the simulated color excess ratios, considering an effective wavelength shift of \ebr=0 and applying the F99 for \rv=3.1 extinction law. Green points represent the color excess ratios from ZY23 for GALEX-related bands.}
\label{sybb}
\end{figure*}

Figure~\ref{sybb} presents the CERs at different \teff\ and compares them with results from other studies. The blue points represent the simulation CERs, accounting for effective wavelength shifts, assuming \ebr=0 and applying the extinction law with F99 for \rv=3.1 . There are noticeable differences in CERs across different \teff, and the trend observed in this study aligns well with the theoretical predictions. For GALEX-related bands, the green points represent the results from ZY23. The results from this work show good agreement with ZY23 for higher-temperature sources. However, significant differences arise when \teff $<$ 5500 K for the NUV band and \teff $<$ 7500 K for the FUV band.

\subsection{Matching the extinction law}

\begin{figure*}
\centering
\includegraphics[width=1\textwidth,angle=0]{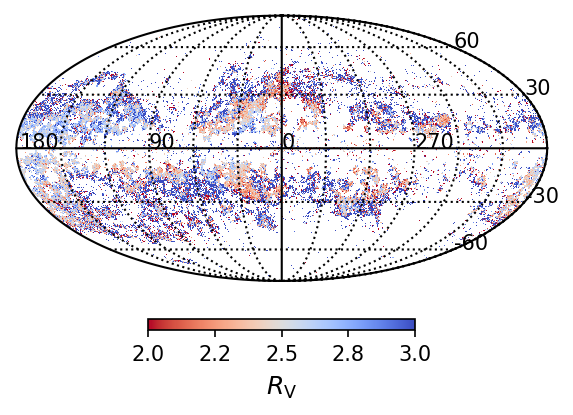}
\caption{The 2D \rv\ map. The colored HEALPix points represent the median \rv\ with \ebv $>$ 0.1 mag for each pixel. The uncolored regions indicate areas with insufficient stars for analysis.}
\label{rvt}
\end{figure*}

The CER can effectively constrain the extinction law and offer insights into dust properties. To investigate the spatial distribution of \rv, the \rv\ value for each star in the sample is first determined. Following a forward modeling approach similar to ZYC23, we derived the best-fit \rv\ for each star by comparing the simulated CERs with the observed values. Specifically, we selected the model closest to the observed \teff\ and \ebv\ (obtained from \ebr\ of this work and $R_{G_{\rm BP},G_{\rm RP}}$ of ZY23). Subsequently, linear interpolation was performed on an \rv\ grid based on the F99 extinction curve to minimize the difference between the observed and simulated CERs.
The sky is divided using the HEALPix scheme ($N_{\text{side}} = 128$), and sources with \ebr\ $>$ 0.1 mag are selected. The median \rv\ value within each pixel is used to represent that region. Figure~\ref{rvt} illustrates the spatial characteristics of \rv\ derived in this work. The \rv\ values cover a broad area within $|b|<50^\circ$ and display clear spatial patterns. In particular, within the CER range covered by this study, \rv\ values are notably lower in regions such as $-30^\circ < l < 30^\circ$ and $20^\circ < b < 50^\circ$, $0^\circ < l < 45^\circ$ and $-30^\circ < b < -10^\circ$, as well as near $60^\circ < l < 160^\circ$ with $b = -15^\circ$. In contrast, relatively higher \rv\ values are found in regions such as $45^\circ < l < 70^\circ$ and $-10^\circ < b < -30^\circ$, near $90^\circ < l < 120^\circ$ with $b = 15^\circ$, and $150^\circ < l < 210^\circ$ with $15^\circ < b < 30^\circ$. In addition, small-scale variations in \rv\ are also observed throughout the entire spatial distribution.

\begin{figure*}
\centering
\includegraphics[width=1\textwidth,angle=0]{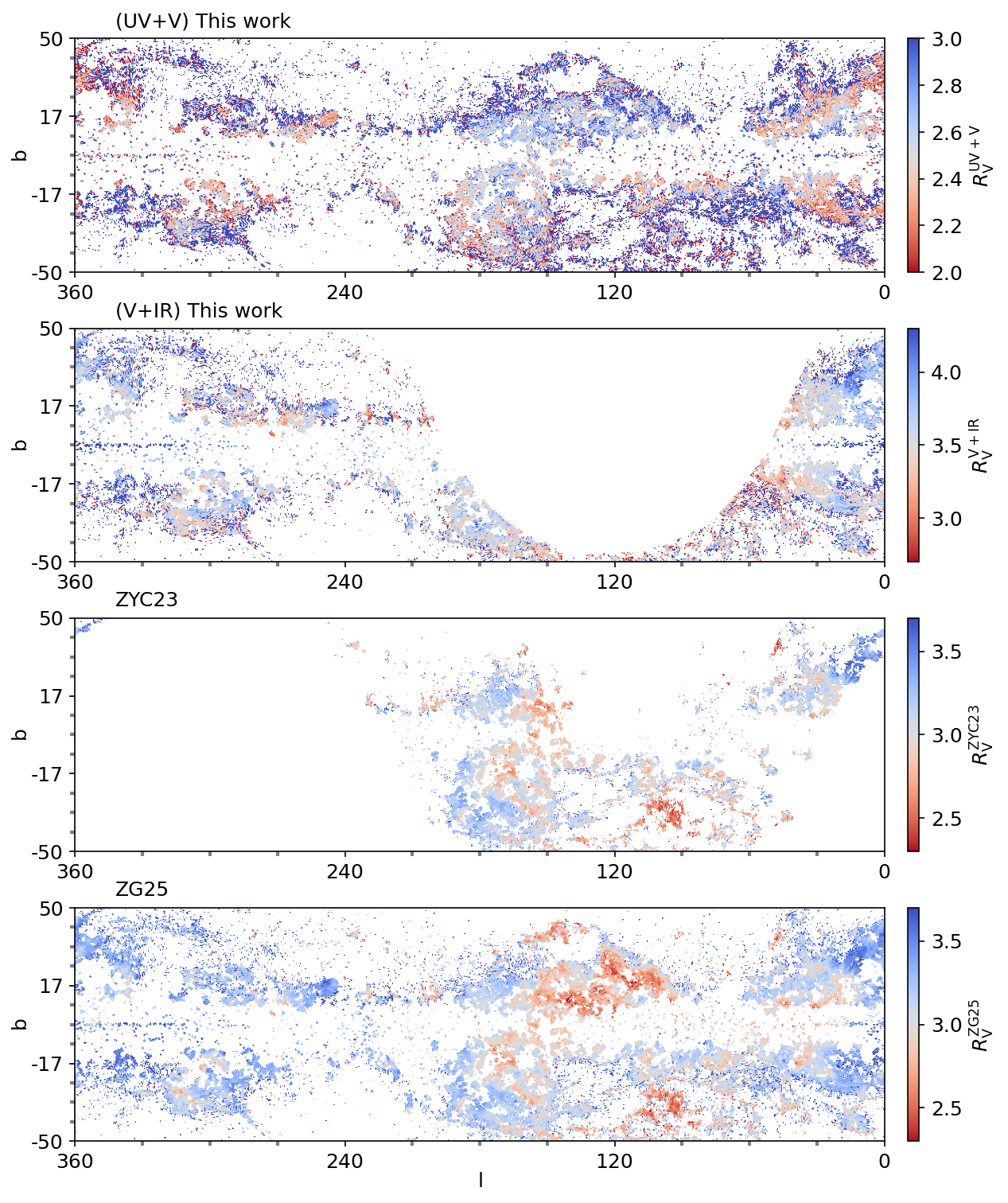}
\caption{Comparison of the two-dimensional \rv\ distribution between the UV common sources of this work and ZYC23, and ZG24. The first panel: the \rv\ distribution ($R_{\rm V}^{\rm UV+V}$) based on UV and Gaia optical band measurements of this work; The second panel: the \rv\ distribution ($R_{\rm V}^{\rm V+IR}$) primarily based on IR and optical bands of this work; The third panel: the \rv\ distribution ($R_{\rm V}^{\rm ZYC23}$) of ZYC23; The fourth panel: the \rv\ distribution ($R_{\rm V}^{\rm ZG25}$) of ZG25.}
\label{rvcm}
\end{figure*}

\begin{figure*}
\centering
\includegraphics[width=1\textwidth,angle=0]{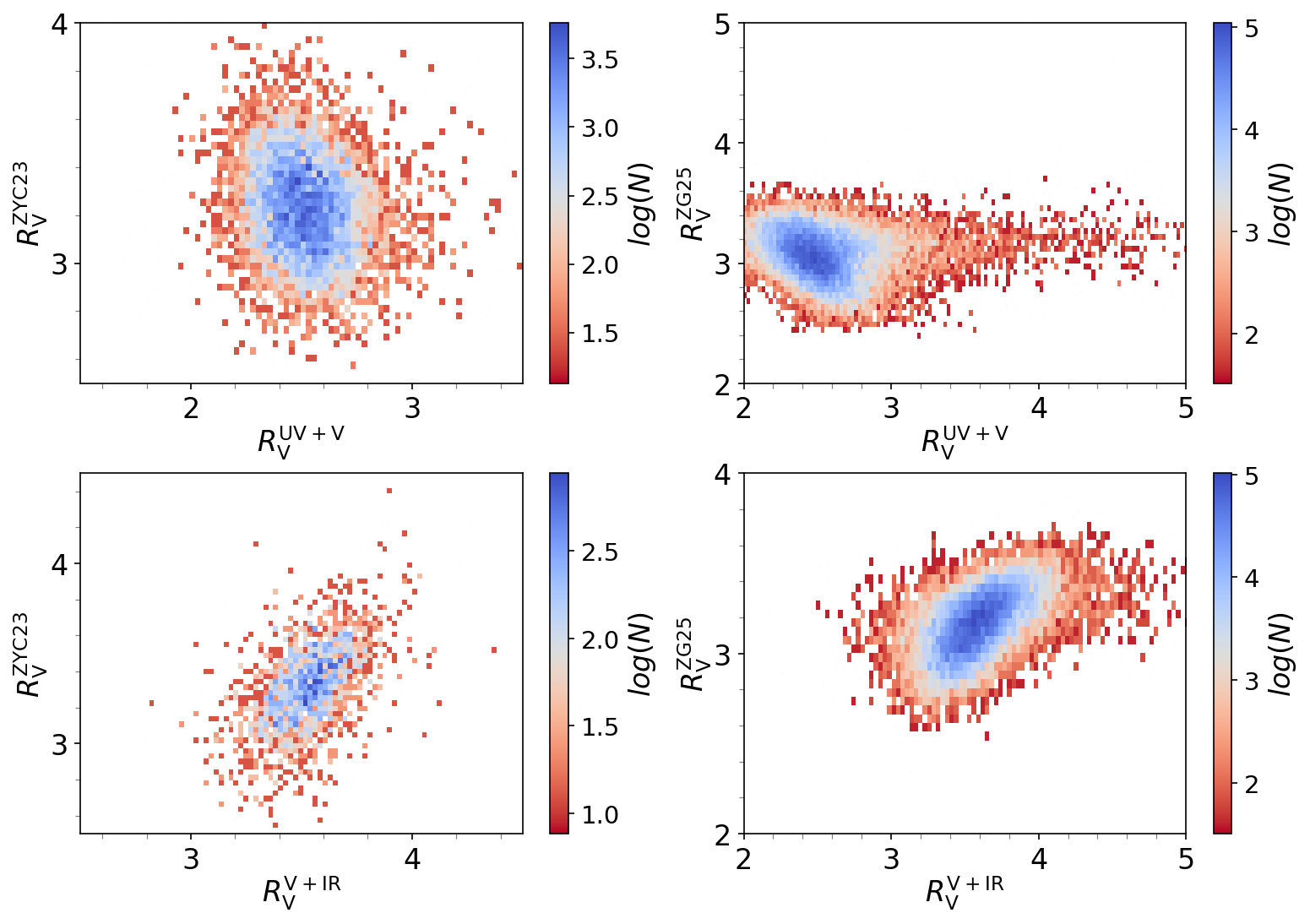}
\caption{Comparison of \rv\ values derived in this work with those from ZYC23 and ZG25 for the same HEALPix pixels. Top panels: Comparison of \rv\ values derived from UV and Gaia optical band measurements in this work with those from ZYC23 (left) and ZG25 (right). Bottom panels: Comparison of \rv\ values primarily derived from IR and optical bands in this work with those from ZYC23 (left) and ZG25 (right). }
\label{rvcm2}
\end{figure*}

The \rv\ distribution derived in this work is compared with those reported by ZYC23 and ZG25, where ZYC23 is based on color excess measurements from mid-infrared to optical bands, and ZG25 utilizes low-resolution Gaia XP spectra. To ensure a fair comparison and eliminate potential biases introduced by selection effects caused by the photometric depth differences between IR/optical and UV bands, the source catalogs from this study, ZYC23, and ZG25 are cross-matched. The common sources identified in the UV band are selected for comparison, as shown in Figure~\ref{rvcm}. It is evident that the \rv\ distribution derived from UV and Gaia optical band measurements in this work (the first panel) differs significantly from the distributions presented by ZYC23 (the third panel) and ZG25 (the fourth panel), which are based on IR/optical data, with some regions even showing an opposite trend. To further clarify these discrepancies and to validate the robustness of our method, we combined multi-band photometric data (Gaia DR3 $G$\bp\rp, WISE $W1W2$, SMSS DR4 $uvgriz$, 2MASS $JH$\ks) with stellar parameters from Z23, applying our blue-edge method to derive color excess measurements. For the UV bands common sources, we produced an \rv\ distribution primarily based on the IR/optical bands using an approach similar to ZYC23 (the second panel), which shows good agreement with both ZYC23 and ZG25. Figure~\ref{rvcm2} compares the linear relationship between the \rv\ values from this work and those from ZYC23 and ZG25 for the same HEALPix pixels. The \rv\ values derived from UV and Gaia optical band measurements in this work show a noticeable anti-correlation with those from ZYC23 and ZG25. In contrast, the \rv\ values primarily derived from infrared/optical bands exhibit a strong positive correlation with those from ZYC23 and ZG25. These results suggest that the UV extinction law is more complex, leading to \rv\ distributions that differ from those of the IR/optical bands. This implies that a single-parameter extinction law, characterized by \rv, may be insufficient to simultaneously characterize extinction behavior across IR, optical and UV bands. Due to space limitations, a deeper discussion of this issue is beyond the scope of this paper. In future work, we will explore the integration of multi-band color excess data to gain a deeper understanding of these differences and to develop improved models for the UV extinction law.

\section{Summary} 
Combining the stellar parameters (\teff, \feh, \logg) provided by L11, G3, A17 and Z23 with photometric data of UV band (NUV, FUV from GALEX and \uvwa, \uvmb, \uvwb\ from UVOT) and optical band (\bp and \rp from Gaia), this work divides the data into small intervals according to stellar parameters. Using the blue edge method and Random Forest techniques, we construct a robust regression model to determine the relationship between stellar parameters and intrinsic color indices in the optical and UV bands. By combining observed color indices, this study obtains the color excesses in the optical, NUV, FUV, \uvwa, \uvmb, and \uvwb\ bands for 92,142,820, 11,624,802, 65,531, 336,633, 137,739, and 253,271 sources, respectively, with typical errors of 0.01, 0.21, 0.30, 0.19, 0.24, and 0.21. We also used the HEALPix method to construct extinction maps in the ultraviolet band, with the NUV extinction map covering about two-thirds of the sky, including most high galactic latitude regions. At the same distance scales, the NUV extinction map from this study exhibits excellent agreement with the large-scale structures of the G19 extinction map and effectively captures the dust structural features in the low-extinction regions at high Galactic latitudes. Based on the UV color excess information derived in this work, we have constructed an \rv\ distribution map primarily covering the $|b|<50^{\circ}$ region. By comparing the \rv\ spatial distribution based on IR/optical bands with those from ZYC23 and ZG25, it is evident that the extinction law in the UV band exhibits significantly higher complexity. The \rv\ distribution in the UV band shows substantial differences from that in the IR/optical bands, with some regions even displaying opposite trends. This result suggests the limitations of a single-parameter \rv\ extinction law in simultaneously characterizing IR, optical, and UV bands.

\section*{Acknowledgements}
We would like to thank the referee for providing us with detailed and constructive feedback that has significantly enhanced the quality of the manuscript. We thank Zhetai Cao for helpful discussions. This work is supported by NSFC through projects 12203016, 12173013, and 12133002, Natural Science Foundation of Hebei Province No.~A2022205018, A2021205006, 226Z7604G, and Science Foundation of Hebei Normal University No.~L2022B33. W.Y.C. acknowledge the support form the science research grants from the China Manned Space Project. M.X.S. acknowledge the support of Physics Postdoctoral Research Station at Hebei Normal University. This work made use of the data taken by \emph{GALEX}, UVOT, \emph{Gaia}.

\facilities{\emph{GALEX}, UVOT, \emph{Gaia}}

\bibliography{UV}{}
\bibliographystyle{aasjournal}

\newpage

\end{CJK*}
\end{document}